\PassOptionsToPackage{dvipsnames}{xcolor}

\documentclass[reprint,superscriptaddress,amsmath,amssymb,longbibliography]{revtex4-2}

\usepackage{hyperref} 

\usepackage{tikz}
\usetikzlibrary{math}

\usepackage{url}
\usepackage{seqsplit}
\usepackage{xstring}
\usepackage[utf8]{inputenc}
\usepackage{graphicx}

\usepackage[caption=false]{subfig} \usepackage{ragged2e} \DeclareCaptionJustification{justified}{\justifying}

\newcommand{\tr}{\operatorname{tr}}

\def\bra#1{\mathinner{\langle{#1}|}}
\def\ket#1{\mathinner{|{#1}\rangle}}

\def\avg#1{\mathinner{\langle{#1} \rangle}}

\begin{document}

\title{Effective quantum volume, fidelity and computational cost of noisy quantum processing experiments}
\author{K. Kechedzhi}\affiliation{Google Quantum AI, California, USA}
\author{S. V. Isakov}\affiliation{Google Quantum AI, Switzerland}
\author{S. Mandr\`a}
\affiliation{Google Quantum AI, California, USA}
\affiliation{Quantum Artificial Intelligence Laboratory, NASA Ames Research Center, Moffett Field, California 94035, USA}
\affiliation{KBR, 601 Jefferson St., Houston, TX 77002, USA}
\author{B. Villalonga}\affiliation{Google Quantum AI, California, USA}
\author{X. Mi}\affiliation{Google Quantum AI, California, USA}
\author{S. Boixo}\affiliation{Google Quantum AI, California, USA}
\author{V. Smelyanskiy}\affiliation{Google Quantum AI, California, USA}
\begin{abstract}
  Today’s experimental noisy quantum processors can compete with and surpass
  all known algorithms on state-of-the-art supercomputers for the computational
  benchmark task of Random Circuit Sampling~\cite{boixo2018characterizing,
  arute2019quantum, wu2021, zhu2022, morvan2023phase}.  Additionally, a
  circuit-based quantum simulation of quantum information
  scrambling~\cite{mi2021information}, which measures a local observable, has
  already outperformed standard full wave function simulation algorithms, e.g.,
  exact Schrodinger evolution and Matrix Product States (MPS). However, this
  experiment has not yet surpassed tensor network contraction for computing the
  value of the observable. Based on those studies, we provide a unified
  framework that utilizes the underlying effective circuit volume to  explain
  the tradeoff between the experimentally achievable signal-to-noise ratio for
  a specific observable, and the corresponding computational cost. We apply
  this framework to recent quantum processor experiments of Random Circuit
  Sampling~\cite{morvan2023phase}, quantum information
  scrambling~\cite{mi2021information}, and a Floquet circuit
  unitary~\cite{kim2023evidence}. This allows us to reproduce the results of
  Ref.~\cite{kim2023evidence} in less than one second per data point using one
  GPU. 
\end{abstract}
\date{June 2023}

\maketitle

\section{Introduction}

Error corrected quantum computers~\cite{feynman2018simulating, shor1995scheme,
gottesman1997stabilizer, knill1998resilient, kitaev2003fault,
dennis2002topological, raussendorf2007fault, fowler2012surface,
satzinger2021realizing, krinner2022realizing, zhao2022realization,
google2023suppressing} have the potential to solve problems that are
intractable for classical computers, including applications in
factoring~\cite{shor1994algorithms, gidney2021factor}, quantum
simulation~\cite{lloyd1996universal} of quantum
chemistry~\cite{aspuru2005simulated, google2020hartree, huggins2022unbiasing,
goings2022reliably} and materials~\cite{rubin2023faulttolerant}, quantum
machine learning~\cite{huang2022quantum} and topological data
analysis~\cite{lloyd2016quantum, berry2022quantifying}. Nevertheless, there are
still many challenges that need to be overcome before full scale error
corrected quantum computers can be realized, most notably reducing the errors
of all hardware components. 

Despite their noise, current experimental quantum processors are already able
to perform significant scientific experiments~\cite{mi2021information,
huggins2022unbiasing, huang2022quantum, frey2022realization, chen2022error,
mi2022time, mi2022noise, morvan2022formation, google2023non, mi2023stable}, and
already surpassed standard  full wave function (``brute force'') simulation
algorithms~\cite{mi2021information}, see also~\footnote{A circuit based quantum
simulation of quantum information scrambling~\cite{mi2021information}, which
measures an expectation value of a  local observable, has surpassed standard
full wave function simulation algorithms, e.g., exact Schrodinger evolution and
Matrix Product States (MPS). We note that this value can still be calculated
with tensor network contraction~\cite{mi2021information}}.

One of the main goals in experimental quantum computing is to demonstrate
computations or simulations with high classical computational cost that address
questions of scientific or commercial interest. Therefore, it has become
essential to understand how to properly gauge the computational cost of a given
quantum processing experiment by generic classical algorithmic methods, as well
as its critical dependence on noise and the specific choice of experimental
observable. 

This type of study has been carried out for Random Circuit Sampling (RCS)
experiments~\cite{boixo2018characterizing, arute2019quantum, wu2021, zhu2022,
morvan2023phase, markov2018quantum, bouland2019complexity,
zlokapa2020boundaries, bouland2022noise, kondo2022quantum,
aaronson2023certified, bassirian2021certified, liu2021benchmarking,
dalzell2022, gao2021limitations, aharonov2022polynomial}.  Those experiments
have long outperformed full wave function simulation
methods~\cite{boixo2018characterizing, de2019massively, pednault2017breaking,
markov2018quantum, arute2019quantum, morvan2022formation}.  There it was found
that tensor network contraction methods~\cite{markov2008simulating,
boixo2017simulation, aaronson2017complexity} offer the most performant generic
classical algorithms to estimate the equivalent classical computational cost,
as first implemented in~\cite{boixo2017simulation}, and significantly improved
by subsequent work~\cite{chen2018classical,
villalonga2019flexible,Villalonga_2020, gray2021hyper, huang2020classical,
kalachev2021multi, villalonga2021efficient, pan2022solving, liu2022validating}.
This framework was first extended to the measurement of a local observable in
Ref.~\cite{mi2021information}.

In this paper we generalize and clarify the tradeoff between a sensitivity to
noise of an  experimental observable, and the corresponding equivalent
classical computational cost to compute the observable.  Consider an operator
$O$ with ideal expectation value $\avg{O}_{\rm ideal}$ and experimental
expectation $\tr(\rho O)$, where $\rho$ is the density matrix of a noisy
quantum state. As we will show, under appropriate experimental conditions, we
often have
\begin{align}
    \tr(\rho O) = F_{\rm eff} \avg{O}_{\rm ideal}\;,
\end{align}
where $F_{\rm eff}$ is the effective fidelity of the observable. The effective fidelity is expected to decay exponentially as 
\begin{align}
    F_{\rm eff} \sim e^{-\epsilon V_{\rm eff}}\;,
\end{align}
where $\epsilon$ is the dominant error per two-qubit entangling gate $\epsilon$
and $V_{\rm eff}$  the effective circuit volume.  For systems without
conservation laws, the effective volume $V_{\rm eff}$ is the number of
entangling two-qubit gates that contribute to the expectation value
$\avg{O}_{\rm ideal}$. Note that $V_{\rm eff}$ depends on the specific
observable $O$. The corresponding computational cost with tensor network
contraction will be exponential in some ``effective area'' $A_{\rm eff}$ of an
appropriate ``cut'' of the effective volume $V_{\rm eff}$, that is,
\begin{align}
    {\rm cost} \propto 2^{\alpha A_{\rm eff}}\;,\label{eq:aeff}
\end{align}
for some constant $\alpha$~\footnote{A more formal definition of the
computational cost of a tensor network contraction shows that it is exponential
in the so called treewidth of the corresponding tensor network.}. (See
Fig.~\ref{fig:a_eff_cartoon} for a cartoon of $V_{\rm eff}$ and $A_{\rm eff}$
for a single qubit Pauli operator.) Therefore, we typically expect a tradeoff
between the desirable high signal-to-noise ratio, and the coveted high
classical computational cost (high $A_{\rm eff})$. 

In Sec.~\ref{sec:rcs} we detail this framework in the context of RCS, where it
arises more naturally. In Sec.~\ref{sec:otoc} we summarize how this framework
was extended to local observables in Ref.~\cite{mi2021information}. In
Sec.~\ref{sec:less} we apply this framework to the Floquet circuit unitary of
Ref.~\cite{kim2023evidence}, and present simulations of this experiment with a
cost of less than one second per data point using one GPU, employing generic
methods. 

\section{Random Circuit Sampling}\label{sec:rcs}

Random quantum circuit sampling (RCS) is an experiment designed to benchmark
both the achievable equivalent computational cost and the full system
fidelity~\cite{boixo2018characterizing,arute2019quantum,morvan2023phase}. In
this case the quantum circuit $C$ is chosen at random from an ensemble of
quantum gates designed to maximize the spread of quantum correlations and the
equivalent classical computational cost. Not coincidentally, this experiment is
also particularly sensitive to noise.

In the presence of noise, the experimental value of an observable $O$ for a
random circuit is
\begin{align}
    \tr(\rho\; O)=F \avg{O}_{\rm ideal} +\frac{1-F}{2^n} \tr(O)\;,
\end{align}
where $\rho$ is the density matrix of a noisy quantum state, $n$ is the number
of qubits and $F$ is the fidelity of the quantum circuit.
The fidelity $F$ can be approximated by 
\begin{align}
    F\simeq e^{-\epsilon V}\;,\label{eq:fid_rcs}
\end{align}
where $V$ is the circuit volume, the number of entangling two-qubit gates, and
$\epsilon$ is the dominant error per gate. 

There is a statistical error $\delta$ when measuring an observable $O$, and we
require $\delta < \tr(\rho\; O)$. For a traceless observable $O$, this puts a
bound on the circuit volume
\begin{align}
V \le \frac 1 \epsilon \log\left({|\avg{O}_{\rm ideal}| \over \delta} \right). ~\label{eq:max-circ-volume}
\end{align}

The corresponding classical  computational cost is estimated by benchmarking an
equivalent classical sampling of bitstrings such that $\avg{O}_{\rm classical}
\simeq F$, using the best classical algorithms available. 

Standard full wave–function quantum circuit simulation algorithms store a
direct representation of the full wave function where gates are applied.
Therefore, despite significant recent improvements, their cost becomes
prohibitive for simulating current quantum processors with around 50 qubits or
more. Similarly, standard full wave-function tensor network algorithms, such as
matrix product states (MPS) or 2D tensor networks, attempt to find a more
efficient representation of the full wave–function when there is only limited
entanglement, but they also become prohibitively expensive as the amount of
entanglement in the quantum state grows. 

Recent research has shown that classical tensor contraction algorithms are the
most performant for the most complex recent quantum computation experiments.
Tensor contraction algorithms are a generalization of quantum circuit
simulation algorithms where each gate is interpreted as a tensor with indices
connecting to neighboring gates. They were first implemented precisely to
benchmark the classical computational cost of random circuit
sampling~\cite{boixo2018characterizing,arute2019quantum,morvan2023phase}. 

For simplicity, consider a unitary circuit acting over time $t$ on a square
$n$-qubit array. The cost of tensor contraction scales like,
\begin{align}
    {\rm cost} \propto 2^{\alpha A}\;,\label{eq:rcs_cost}
\end{align}
where $A \sim \min(\sqrt n t, n)$ is the area of a cut of the circuit
volume~\cite{Note2}. For short evolution times $A = \sqrt n t$ and the cost is
exponential in the maximum amount of entanglement generated by the circuit. The
coefficient $\alpha$ depends on the entangling power of the gate used. For
instance, $\alpha = 2$ for iSWAP gates, while $\alpha = 1$ for CZ
gates~\cite{markov2018quantum}, determined by the Schmidt coefficients of the
gate.

Therefore, we arrive at the following intuitive conclusion: for a fixed gate
error $\epsilon$, the feasibility or fidelity of the experiment decreases
exponentially with the circuit volume, while the corresponding classical
computational cost increases exponentially with the area $A$ of a cut of the
circuit volume.  The RCS experiment reported in Ref.~\cite{morvan2023phase} has
an estimated contraction cost larger than $2^{78}$~\cite{footnote:FLOPs} and
fidelity of $1.68 \cdot 10^{-3}$ for the hardest circuits. The average error
per iSWAP-like entangling gate was $\sim 0.67\%$,  and the (effective) volume
was 702 gates. The estimated classical simulation time is 47.2 years if using
the Frontier supercomputer, the only exascale supercomputer in the TOP500 list. 
 
\section{Information scrambling with local operators}\label{sec:otoc}

The relationship between gate error, the fidelity of an observable, and the
corresponding computational cost, has also been analyzed recently in the
context of quantum information scrambling~\footnote{Understanding how quantum
information scrambles is essential to understanding a number of physical
phenomena, such as the fast-scrambling conjecture for black holes, non-Fermi
liquid behaviors, and many-body localization. Ref.~\cite{mi2021information}
also provides a basis for designing near term quantum algorithms that would
benefit from efficient exploration of the Hilbert space. } with
out-of-time-correlator (OTOC) measurements Ref.~\cite{mi2021information}.
Quantum information scrambling can be seen as a ``butterfly effect'', in which
a small change in one location can quickly grow into a large change over time.
More precisely, the disturbance is implemented as an initially local operator
(the ``butterfly operator'') $B$, typically a Pauli operator acting on one of
the qubits (the "butterfly qubit"). Ref.~\cite{mi2021information} investigated
the dynamic process in which the butterfly operator $B$ evolves under a random
quantum circuit $C$ as $B(t) = C^\dagger B C$, where $C^\dagger$ is the inverse
of $C$. The information scrambling of this process is studied with an  OTOC
measurement $\avg{B(t)MB(t)M}$ where $M$ is another Pauli operator on a
different qubit. The OTOC observable is encoded in the Pauli ${\rm Y}_a$
expectation value of an ancilla qubit ``a'' (see Ref.~\cite{mi2021information}
for details). 

The observable $\tr({\rm Y}_a \rho)$ considered in
Ref.~\cite{mi2021information} averaged over random circuits subject to
experimental noise can be written in the first approximation as,  
\begin{align}
    \tr({\rm Y}_a \rho) = F_{\rm eff}\avg{{\rm Y}_a}_{\rm ideal}\;,\label{eq:y_otoc}
\end{align}
where $F_{\rm eff}$ is an effective fidelity~\footnote{See
Ref.~\cite{mi2021information} for better approximations that take into account
experimental bias of the outcome measurements and details of the propagation of
errors for specific circuits.}. The effective fidelity decays exponentially
with the dominant error per two-qubit entangling gate $\epsilon$ and the
effective circuit volume $V_{\rm eff}$ as
\begin{align}
    F_{\rm eff} \sim e^{-\epsilon V_{\rm eff}}\;.\label{eq:F_eff}
\end{align} 
The effective volume $V_{\rm eff}$ is the number of entangling gates that
contribute to the expectation value $\avg{{\rm Y}_a}_{\rm ideal}$. In other
words, removing any gate from $V_{\rm eff}$ would result in a relevant finite
error $\delta\ll1$  in the reduced value of the observable $\avg{{\rm
Y}_a}_{\rm reduced}$, i.e. $\left|\avg{{\rm Y}_a}_{\rm ideal} -\avg{{\rm
Y}_a}_{\rm reduced}\right|\geq \delta$. Equation~(\ref{eq:y_otoc}) offers a way
to perform error mitigation by estimating the effective fidelity $F_{\rm eff}$
so that, 
\begin{align}
    \avg{{\rm Y}_a}_{\rm ideal} = {\tr({\rm Y}_a \rho) \over F_{\rm eff}}\;,\label{eq:otoc_em}
\end{align}
as demonstrated in Refs.~\cite{arute2020observation,mi2021information}. 

A single-qubit butterfly operator $B$ evolved with a circuit unitary is
transformed into $B(t) = C^\dagger B C$ which has support on multiple qubits.
This transformation proceeds via operator spreading: every time an entangling
gate is applied to a pair of qubits crossing the boundary of $B(t)$ the support
increases with some probability $p$. One might expect that under operator
spreading $V_{\rm eff}$ is the subset of gates within the respective light cone
of $B$. However, this corresponds only to the case of $p=1$,  realized for
example by a random circuit with iSWAP entangling gates. In generic random
circuits the probability of spreading between two qubits is $p\le1$. The
boundary of the support of $B(t)$ remains relatively sharp, spreading at a
(butterfly) velocity of $v_B \leq v_{LC} = 1$, where $v_{LC}$ is the light cone
velocity.  Reference~\cite{mi2021information} shows that, as an example, the
butterfly velocity for $\sqrt{\rm iSWAP}$ random circuits is indeed less than
the butterfly velocity of iSWAP random circuits. The butterfly velocity
determines the size of the effective volume,
\begin{align}
    V_{\rm eff} \propto \min\left((v_B t)^2 t, nt \right)\;,\label{eq:v_eff_scaling}
\end{align}
up to a factor of order one. In a noisy device only the noise channels
associated with entangling gates within $V_{\rm eff}$ contribute to the
measurement outcome which is reflected in Eqs.~(\ref{eq:y_otoc})
and~(\ref{eq:F_eff}).

We note that the experiment reported in Ref.~\cite{mi2021information} can not
be simulated by standard full wave function simulation algorithms, including
full wave function matrix product states (MPS) or 2D tensor networks (iso-TNS).
Indeed, in these circuits the entanglement spreads fast, and the entanglement
entropy between two equal halves of the quantum system saturates quickly. The
bond dimension required for a full wave function tensor network simulation for
the circuits used in Ref.~\cite{mi2021information} would be $\chi \sim 2^{25}$,
see App.~\ref{app:otoc_entanglement}.

Nevertheless, as explained above, these are not the most efficient simulation
algorithms. Tensor network contraction can be adapted to the problem of
estimating the value of an observable, such as ${\rm Y}_a$. Specifically, we
can perform a tensor contraction where some output indices corresponding to
some qubits are left “open”, i.e., they are not assigned to the value of a
given output bitstring. The corresponding contraction produces a vector which
encodes all the corresponding amplitudes for all possible assignments of the
corresponding bits. The cost of a tensor network contraction with a relatively
small number of open qubits is not much higher than a contraction with all
“closed” or assigned qubits. Furthermore, it can be seen that the value of a
local observable is independent of the value assigned to indices outside a
relatively small neighborhood that contains the observable, see
Ref.~\cite{mi2021information}.  

\begin{figure}[t!]
    \centering
    \includegraphics[width=0.5\textwidth]{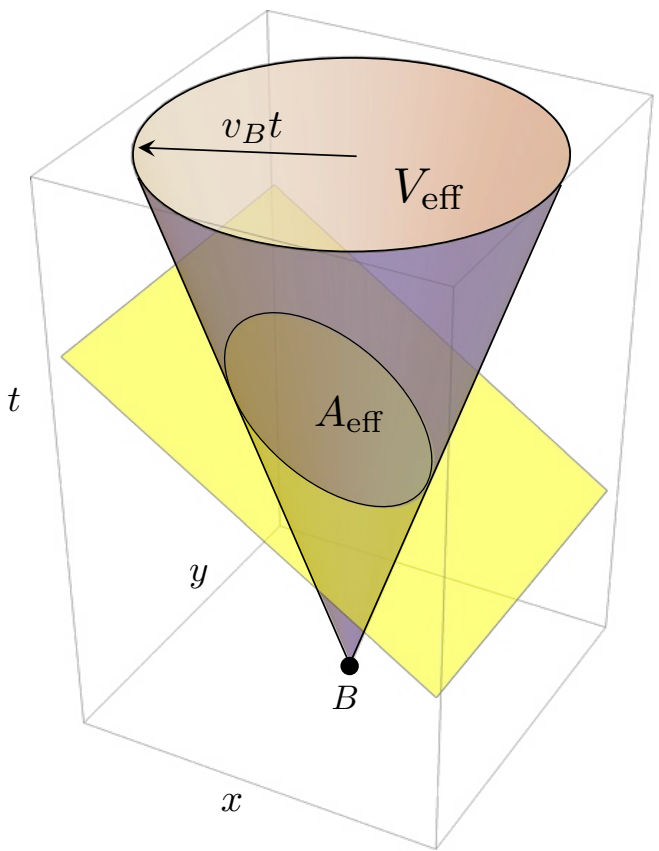}
    \caption{The cartoon depicts the conical surface in $(x,y,t)$ space with
    the volume $V_{\rm eff}$ that envelopes the tensorial structure of
    entangling gates contributing to the local observable $B(t)=U^\dag(t) B
    U(t)$ where $B(0)=B$ is depicted with a black dot. The cut with area
    $A_{\rm eff}$ discussed in the text is represented by the cross-section of
    the plane and the cone. The base of the cone corresponds to the subset of
    qubits in $(x,y)$ plane that are involved in the operator spreading of  $B$
    at time $t$. The perimeter of the base corresponds to the scrambling front
    that moves with the velocity $v_B$.}
    \label{fig:a_eff_cartoon}
\end{figure}

To accurately estimate the cost of an observable for a given circuit
simulation, it is critical to note that the required tensor network only needs
to contain the quantum gates within the effective volume. Therefore, the cost
will be exponential in some ``effective area'' $A_{\rm eff}$ of an appropriate
``cut'' of the effective volume,

as in Eq.~\eqref{eq:rcs_cost}~\footnote{More formally, the cost will be
exponential in the treewidth $A_{\rm eff}$ of the line graph of the tensor
network corresponding to the quantum circuit contained within the effective
volume $V_{\rm eff}$}, see Fig.~\ref{fig:a_eff_cartoon}. Note that we assume as
before that the circuit acts in a two dimensional array of qubits.  We also
have 
\begin{align}
    A_{\rm eff} \le \min\left((v_B t)^2,n\right)\;,\label{eq:a_eff_scaling}
\end{align}
with Butterfly velocity $v_B$. For a long one dimensional array of qubits
$A_{\rm eff} \le \min\left(2 v_B t,n\right)$. Note the difference between
Eqs.~(\ref{eq:aeff}) and~(\ref{eq:a_eff_scaling}) is due to locality of
operators $M$ and $B$ whereas in Eqs.~(\ref{eq:aeff}) the measured operator has
support on all qubits.

We arrive at a conclusion similar to the one for random circuit sampling.
Namely, an experiment in a quantum processor is fundamentally limited by the
exponential decrease in fidelity with the effective quantum volume, while the
equivalent classical computational cost increases exponentially with an
effective area $A_{\rm eff}$. In other words, while experimentally we would
like a small effective volume to obtain a good experimental signal-to-noise
ratio, this consideration will limit the corresponding achievable computational
complexity. We emphasize that the exponent in the fidelity, $V_{\rm eff}$, and
the exponent in the cost $A_{\rm eff}$, have different scaling, see
Eqs.~\eqref{eq:v_eff_scaling} and \eqref{eq:a_eff_scaling}.

The OTOC experiment reported in Ref.~\cite{mi2021information} had an effective
fidelity $F_{\rm eff} \simeq 0.06$, with an effective volume (light cone of
iSWAPs) of 251 gates for the circuit size where the error mitigated
signal-to-noise ratio was 1. The contraction cost was $2^{41.2}$. The largest
experiment reported in that reference, with a smaller signal-to-noise ratio,
had an effective fidelity $F_{\rm eff} \simeq 0.02$, with an effective volume
of $292$ iSWAP gates and a contraction cost $2^{49.3}$. The average error per
iSWAP gate was $\sim 0.9\%$.

As discussed above the cost estimated in Eq.~(\ref{eq:a_eff_scaling})
corresponds to a general purpose tensor network contraction algorithm. However,
often there exist model-specific approximate methods that use analytical
insights to significantly reduce the computational cost, such as stabilizer
based algorithms exploiting the Clifford group, or mappings to free fermions.
Another instructive example was discussed in Ref.~\cite{mi2021information},
where the value of OTOC averaged over instances of random circuits can be
predicted at a fraction of the computational cost, specifically using a mapping
onto a classical Markov chain. At the same time, there are no known efficient
classical methods to compute the value of OTOC for a specific instance of a
random circuit, and tensor contraction remains the best performing algorithm.
The effective circuit volume of such a simulation is estimated from the
precision needed to resolve the circuit-specific variation of OTOC.  Therefore,
the variance of OTOC determines the conditions on the above task for a quantum
simulation to surpass classical supercomputers for all known classical
algorithms. 
 
\section{Time ordered observables}\label{sec:less}

The above considerations apply to experiments in quantum processors more
generally. We consider for simplicity experiments  estimating a (time ordered)
expectation value of a string of Pauli operators $P_S$ with support in $S$
qubits. Under appropriate experimental conditions we expect, as in
Eq.~\eqref{eq:y_otoc},
\begin{align}
       \tr(P_S \rho) = F_{\rm eff}\avg{P    _S}_{\rm ideal}\;,\label{eq:P_S}
\end{align}
where $F_{\rm eff} \sim e^{-\epsilon V_{\rm eff}}$ is the effective fidelity of
the experimental measurement and $V_{\rm eff}$ is the corresponding effective
circuit volume, see Eq.~\eqref{eq:F_eff}. 

We first consider the evolution of a local operator in the generic case when
the quantum state becomes chaotic (or ergodic) and the information scrambling
takes hold. In this case $\avg{P_S}_{\rm ideal}$ decays very rapidly to 0 on a
time scale given by the inverse of the butterfly velocity $1/v_B$. The
corresponding effective volume and the cost will be limited by statistical
error or by the effective fidelity, see App.~\ref{app:chaotic}. 

A more interesting case is when the transition to chaos is preceded at early
stages by a so-called prethermalization regime where $\avg{P_S}_{\rm ideal}$
can decay slowly to 0. One example is given by Floquet circuits with layers of
control-phase gates and single qubit $X$ rotations with not too large angles.
This typically corresponds to a Floquet transverse field Ising model, where the
cycle unitary and the corresponding Floquet Hamiltonian posses an Ising
symmetry leading to a spontaneous symmetry broken phase away from the critical
point. This generic model has been used in experimental simulations exploring
quantum many-body phenomena, such as time
crystals~\cite{mi2022time,frey2022realization}, quantum
scars~\cite{chen2022error}, Majorana edge modes~\cite{mi2022noise} and
dissipative bath engineering~\cite{mi2023stable}. 

A recent experiment with this system consists of $\pi/2$ ZZ rotations and X
rotations with the same angle $\theta_h \in [0,\pi/2]$ for all the
qubits~\cite{kim2023evidence}. Explicitly, the Floquet circuit is
\begin{align}
    \prod_{\langle j,k \rangle}\exp \left({\rm i} \frac{\pi}{4}Z_{j}Z_{k}
    \right)\,\prod_{j}\exp \left(-{\rm i}
    \frac{\theta_{h}}{2}X_{j}\right)\;,\label{eq:floquet}
\end{align}
where $\langle j,k \rangle$ denotes nearest neighbors.  The initial condition,
all spins up, corresponds, with small angles $\theta_h$, to a system deeply in
the ferromagnetic symmetry broken phase. There the $\avg{Z_j}$ magnetization is
large and the $\avg{Z_j Z_k} - \avg{Z_j} \avg{Z_k}$ correlators decay quickly
with distance (short correlation length). 

The experiment~\cite{kim2023evidence} uses an error mitigation technique that
similarly to the one explained in Sec.~\ref{sec:otoc}, relies on
Eq.~\eqref{eq:otoc_em}. The effective fidelity $F_{\rm eff}$ is estimated
experimentally by adding errors in a controlled way, which is related to
increasing the component error $\epsilon$, and fitting the corresponding
exponential decay of $F_{\rm eff}$. 

\subsection{Local time-ordered observable}\label{sec:local_ferro}

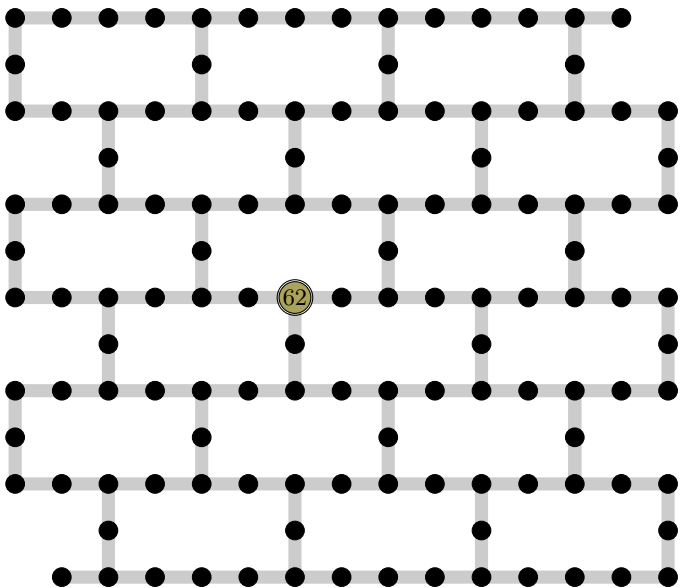
\begin{figure}
    \centering
    
\def\scale{0.62}
\def\circlesize{0.2}
\def\vspace{1.0}
\begin{tikzpicture}
\foreach \x [evaluate=\x as \y using \x + 1] in {1, 2, 3, 4, 5, 6, 7, 8, 9, 10, 11, 12, 13}
    \draw[black!20!white, line width=8*\scale] (\x*\scale, 13*\vspace*\scale) -- (\y*\scale, 13*\vspace*\scale);
  \foreach \x [evaluate=\x as \y using \x + 1] in {1, 2, 3, 4, 5, 6, 7, 8, 9, 10, 11, 12, 13, 14}
    \draw[black!20!white, line width=8*\scale] (\x*\scale, 11*\vspace*\scale) -- (\y*\scale, 11*\vspace*\scale);
  \foreach \x [evaluate=\x as \y using \x + 1] in {1, 2, 3, 4, 5, 6, 7, 8, 9, 10, 11, 12, 13, 14}
    \draw[black!20!white, line width=8*\scale] (\x*\scale, 9*\vspace*\scale) -- (\y*\scale, 9*\vspace*\scale);
  \foreach \x [evaluate=\x as \y using \x + 1] in {1, 2, 3, 4, 5, 6, 7, 8, 9, 10, 11, 12, 13, 14}
    \draw[black!20!white, line width=8*\scale] (\x*\scale, 7*\vspace*\scale) -- (\y*\scale, 7*\vspace*\scale);
  \foreach \x [evaluate=\x as \y using \x + 1] in {1, 2, 3, 4, 5, 6, 7, 8, 9, 10, 11, 12, 13, 14}
    \draw[black!20!white, line width=8*\scale] (\x*\scale, 5*\vspace*\scale) -- (\y*\scale, 5*\vspace*\scale);
  \foreach \x [evaluate=\x as \y using \x + 1] in {1, 2, 3, 4, 5, 6, 7, 8, 9, 10, 11, 12, 13, 14}
    \draw[black!20!white, line width=8*\scale] (\x*\scale, 3*\vspace*\scale) -- (\y*\scale, 3*\vspace*\scale);
  \foreach \x [evaluate=\x as \y using \x + 1] in {2, 3, 4, 5, 6, 7, 8, 9, 10, 11, 12, 13, 14}
    \draw[black!20!white, line width=8*\scale] (\x*\scale, 1*\vspace*\scale) -- (\y*\scale, 1*\vspace*\scale);

\foreach \y [evaluate=\y as \z using \y - 1] in {13, 12, 9, 8, 5, 4}
    \draw[black!20!white, line width=8*\scale] (1*\scale, \y*\vspace*\scale) -- (1*\scale, \z*\vspace*\scale);
  \foreach \y [evaluate=\y as \z using \y - 1] in {13, 12, 9, 8, 5, 4}
    \draw[black!20!white, line width=8*\scale] (5*\scale, \y*\vspace*\scale) -- (5*\scale, \z*\vspace*\scale);
  \foreach \y [evaluate=\y as \z using \y - 1] in {13, 12, 9, 8, 5, 4}
    \draw[black!20!white, line width=8*\scale] (9*\scale, \y*\vspace*\scale) -- (9*\scale, \z*\vspace*\scale);
  \foreach \y [evaluate=\y as \z using \y - 1] in {13, 12, 9, 8, 5, 4}
    \draw[black!20!white, line width=8*\scale] (13*\scale, \y*\vspace*\scale) -- (13*\scale, \z*\vspace*\scale);
  \foreach \y [evaluate=\y as \z using \y - 1] in {13, 12, 9, 8, 5, 4}
    \draw[black!20!white, line width=8*\scale] (13*\scale, \y*\vspace*\scale) -- (13*\scale, \z*\vspace*\scale);
  \foreach \y [evaluate=\y as \z using \y - 1] in {11, 10, 7, 6, 3, 2}
    \draw[black!20!white, line width=8*\scale] (3*\scale, \y*\vspace*\scale) -- (3*\scale, \z*\vspace*\scale);
  \foreach \y [evaluate=\y as \z using \y - 1] in {11, 10, 7, 6, 3, 2}
    \draw[black!20!white, line width=8*\scale] (7*\scale, \y*\vspace*\scale) -- (7*\scale, \z*\vspace*\scale);
  \foreach \y [evaluate=\y as \z using \y - 1] in {11, 10, 7, 6, 3, 2}
    \draw[black!20!white, line width=8*\scale] (11*\scale, \y*\vspace*\scale) -- (11*\scale, \z*\vspace*\scale);
  \foreach \y [evaluate=\y as \z using \y - 1] in {11, 10, 7, 6, 3, 2}
    \draw[black!20!white, line width=8*\scale] (15*\scale, \y*\vspace*\scale) -- (15*\scale, \z*\vspace*\scale);

\foreach \x in {1, 2, 3, 4, 5, 6, 7, 8, 9, 10, 11, 12, 13, 14}
    \filldraw[fill=black, draw=black] (\x*\scale, 13*\vspace*\scale) circle (\circlesize*\scale);
  \foreach \x in {1, 5, 9, 13}
    \filldraw[fill=black, draw=black] (\x*\scale, 12*\vspace*\scale) circle (\circlesize*\scale);
  \foreach \x in {1, 2, 3, 4, 5, 6, 7, 8, 9, 10, 11, 12, 13, 14, 15}
    \filldraw[fill=black, draw=black] (\x*\scale, 11 * \vspace*\scale) circle (\circlesize*\scale);
  \foreach \x in {3, 7, 11, 15}
    \filldraw[fill=black, draw=black] (\x*\scale, 10*\vspace*\scale) circle (\circlesize*\scale);
  \foreach \x in {1, 2, 3, 4, 5, 6, 7, 8, 9, 10, 11, 12, 13, 14, 15}
    \filldraw[fill=black, draw=black] (\x*\scale, 9*\vspace*\scale) circle (\circlesize*\scale);
  \foreach \x in {1, 5, 9, 13}
    \filldraw[fill=black, draw=black] (\x*\scale, 8*\vspace*\scale) circle (\circlesize*\scale);
  \foreach \x in {1, 2, 3, 4, 5, 6, 7, 8, 9, 10, 11, 12, 13, 14, 15}
    \filldraw[fill=black, draw=black] (\x*\scale, 7*\vspace*\scale) circle (\circlesize*\scale);
  \foreach \x in {3, 7, 11, 15}
    \filldraw[fill=black, draw=black] (\x*\scale, 6*\vspace*\scale) circle (\circlesize*\scale);
  \foreach \x in {1, 2, 3, 4, 5, 6, 7, 8, 9, 10, 11, 12, 13, 14, 15}
    \filldraw[fill=black, draw=black] (\x*\scale, 5*\vspace*\scale) circle (\circlesize*\scale);
  \foreach \x in {1, 5, 9, 13}
    \filldraw[fill=black, draw=black] (\x*\scale, 4*\vspace*\scale) circle (\circlesize*\scale);
  \foreach \x in {1, 2, 3, 4, 5, 6, 7, 8, 9, 10, 11, 12, 13, 14, 15}
    \filldraw[fill=black, draw=black] (\x*\scale, 3*\vspace*\scale) circle (\circlesize*\scale);
  \foreach \x in {3, 7, 11, 15}
    \filldraw[fill=black, draw=black] (\x*\scale, 2*\vspace*\scale) circle (\circlesize*\scale);
  \foreach \x in {2, 3, 4, 5, 6, 7, 8, 9, 10, 11, 12, 13, 14, 15}
    \filldraw[fill=black, draw=black] (\x*\scale, 1*\vspace*\scale) circle (\circlesize*\scale);

\tikzmath{
    \ccirclesize = 1.7*\circlesize;
  }
  \filldraw[fill=yellow!60!black, draw=black!60!black] (7*\scale, 7*\vspace*\scale) circle (\ccirclesize*\scale);
  \node[circle, draw, inner sep=0.1em, text=black] at (7*\scale, 7*\vspace*\scale) {62};

\end{tikzpicture}
    \caption{Layout of the 127 qubits used in the device of Ref.~\cite{kim2023evidence}, and the position of the qubit labelled 62.}
    \label{fig:127_device}
\end{figure}

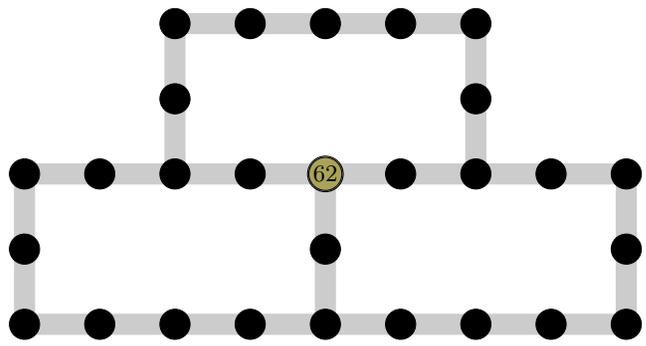
\begin{figure}[h!]
    \centering
    
\def\scale{1.0}
\def\circlesize{0.2}
\def\vspace{1.0}
\begin{tikzpicture}
\foreach \x [evaluate=\x as \y using \x + 1] in {5, 6, 7, 8}
    \draw[black!20!white, line width=8*\scale] (\x*\scale, 5*\vspace*\scale) -- (\y*\scale, 5*\vspace*\scale);
  \foreach \x [evaluate=\x as \y using \x + 1] in {3, 4, 5, 6, 7, 8, 9, 10}
    \draw[black!20!white, line width=8*\scale] (\x*\scale, 3*\vspace*\scale) -- (\y*\scale, 3*\vspace*\scale);
  \foreach \x [evaluate=\x as \y using \x + 1] in {3, 4, 5, 6, 7, 8, 9, 10}
    \draw[black!20!white, line width=8*\scale] (\x*\scale, 1*\vspace*\scale) -- (\y*\scale, 1*\vspace*\scale);

\foreach \y [evaluate=\y as \z using \y - 1] in {5, 4}
    \draw[black!20!white, line width=8*\scale] (5*\scale, \y*\vspace*\scale) -- (5*\scale, \z*\vspace*\scale);
  \foreach \y [evaluate=\y as \z using \y - 1] in {5, 4}
    \draw[black!20!white, line width=8*\scale] (9*\scale, \y*\vspace*\scale) -- (9*\scale, \z*\vspace*\scale);
  \foreach \y [evaluate=\y as \z using \y - 1] in {3, 2}
    \draw[black!20!white, line width=8*\scale] (3*\scale, \y*\vspace*\scale) -- (3*\scale, \z*\vspace*\scale);
  \foreach \y [evaluate=\y as \z using \y - 1] in {3, 2}
    \draw[black!20!white, line width=8*\scale] (7*\scale, \y*\vspace*\scale) -- (7*\scale, \z*\vspace*\scale);
  \foreach \y [evaluate=\y as \z using \y - 1] in {3, 2}
    \draw[black!20!white, line width=8*\scale] (11*\scale, \y*\vspace*\scale) -- (11*\scale, \z*\vspace*\scale);

\foreach \x in {5, 6, 7, 8, 9}
    \filldraw[fill=black, draw=black] (\x*\scale, 5*\vspace*\scale) circle (\circlesize*\scale);
  \foreach \x in {5, 9}
    \filldraw[fill=black, draw=black] (\x*\scale, 4*\vspace*\scale) circle (\circlesize*\scale);
  \foreach \x in {3, 4, 5, 6, 7, 8, 9, 10, 11}
    \filldraw[fill=black, draw=black] (\x*\scale, 3*\vspace*\scale) circle (\circlesize*\scale);
  \foreach \x in {3, 7, 11}
    \filldraw[fill=black, draw=black] (\x*\scale, 2*\vspace*\scale) circle (\circlesize*\scale);
  \foreach \x in {3, 4, 5, 6, 7, 8, 9, 10, 11}
    \filldraw[fill=black, draw=black] (\x*\scale, 1*\vspace*\scale) circle (\circlesize*\scale);

\tikzmath{
    \ccirclesize = 1.1*\circlesize;
  }
  \filldraw[fill=yellow!60!black, draw=black!60!black] (7*\scale, 3*\vspace*\scale) circle (\ccirclesize*\scale);
  \node[circle, draw, inner sep=0.1em, text=black] at (7*\scale, 3*\vspace*\scale) {62};

\end{tikzpicture}

    \caption{Qubits layout around the qubit labelled 62 (in yellow) in Ref.~\cite{kim2023evidence}.}
    \label{fig:q62}
\end{figure}

In some of the experiments reported in Ref.~\cite{kim2023evidence}, including
the ones corresponding to the largest circuits, the measurement consists of a
single $Z$ Pauli operator as a function of $\theta_h$. One way to estimate the
effective circuit volume for a given observable from experimental data is by
using the corresponding effective fidelity. This can be determined from the
ratio of unmitigated and mitigated data points in the experimental data shown
in Fig.~4b of Ref.~\cite{kim2023evidence}, see also Fig.~\ref{fig:q62}. The
data shows $\avg{Z}$ for the qubit labeled 62 with 20 steps (60 layers of
two-qubit gates) of the Floquet circuit of Eq.~\eqref{eq:floquet}. The light
cone covers all 127 qubits in Fig.~\ref{fig:127_device}. For the data points
corresponding to $\theta_h =\pi/4$ we obtain  
\begin{align}
    {\avg{Z_{62}}_{\rm unmitigated}\over \avg{Z_{62}}_{\rm mitigated}} \approx 0.37\;,
\end{align} 
and lager values for $\theta_h<\pi/4$. Using the average error rate $0.01$ per
entangling gate reported in Ref.~\cite{kim2023evidence} we obtain an estimate
of the effective circuit volume of $V_{\rm eff} \sim 100$ gates. Note that
$V_{\rm eff}$ is substantially smaller than the 2880 two-qubit entangling gates
in the circuit. The fidelity of a circuit for this effective volume would be
$\sim 10^{-12}$. This suggests that the cost of an optimal tensor network
contraction for such a local observable must be significantly lower than
simulating state evolution subject to all of the gates applied in the
experiment.

\begin{figure}[t!]
    \centering
    \includegraphics[width=0.5\textwidth]{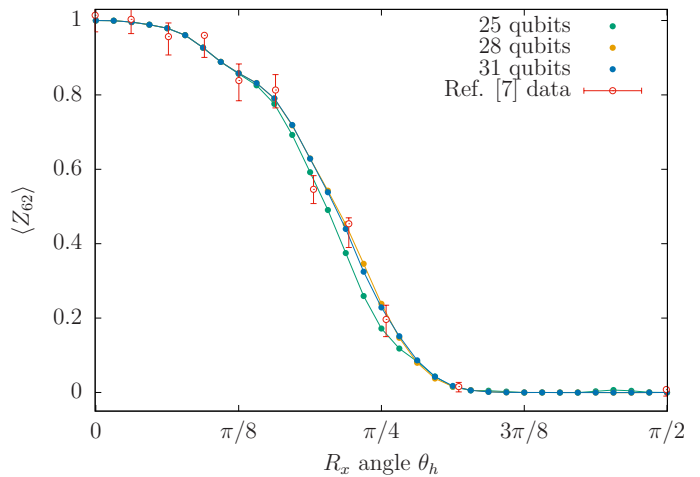}
    \caption{Numerical simulations for $\avg{Z}$ of the qubit labelled 62 in
    Fig.~\ref{fig:q62}, with 20 steps of the Floquet circuit of
    Eq.~\eqref{eq:floquet}. Figure also shows the experimental data reported in
    Fig.~4b of Ref.~\cite{kim2023evidence}. The difference between numerical
    simulations with 28 and 31 qubits is much smaller than the 1-sigma error
    bars of the experiment for all values of $\theta_h$. The simulation with 28
    qubits includes the three loops in Fig.~\ref{fig:q62}. Each point per
    $\theta_h$ value is obtained in less than one second on an A100 GPU using
    the open source simulator
    qsim~\cite{quantum_ai_team_and_collaborators_2020_4023103,
    isakov2021simulations, footnote:code}.}\label{fig:fig4b}
\end{figure}

This suggests that a simulation of a smaller circuit could reproduce the
observed data points. Indeed in Fig.~\ref{fig:fig4b} we compare experimental
data points extracted from Fig.~4b of Ref.~\cite{kim2023evidence} to numerical
simulations with a smaller number of qubits $n\in\{25,28,31\}$. The difference
between numerical simulations with 28 and 31 qubits is much smaller than the
1-sigma error bars of the experiment for all values of $\theta_h$. This good
agreement, the clear convergence seen at lower depth (see
App.~\ref{app:convergence}), and related numerics in
App.~\ref{app:closed_vs_open}, suggest that the numerical simulation is more
accurate than the experiment. See also App.~\ref{app:convergence} for the
convergence of the magnetization in a smaller experiment, where it again does
not require the simulation of the full light cone. Simulations with 25 qubits
or fewer show good convergence for $\theta_h \lesssim \pi/8$, corresponding to
the small correlation length deeply in the ferromagnetic symmetry broken phase.
We also see a good agreement for large $\theta_h \gtrsim 3 \pi/8$,
corresponding to the ergodic case. Each point per $\theta_h$ value is obtained
in 0.8 seconds on an A100 GPU using the open source simulator
qsim~\cite{quantum_ai_team_and_collaborators_2020_4023103,isakov2021simulations,
footnote:code} for $n=28$, and 6.9 seconds per $\theta_h$ for
$n=31$~\footnote{In the process of finishing this paper, another paper appeared
with a different method that performs the same
simulations~\cite{tindall2023efficient}. Our results and the results in this
reference agree, except for large $\theta_h$ values reported in Fig. 3b in
Ref.~\cite{tindall2023efficient}, Fig. 4b in Ref.~\cite{kim2023evidence} and
Figs.~\ref{fig:fig4b} and~\ref{fig:fig4b2} here. We note that
Ref.~\cite{tindall2023efficient} overestimates the value of $\avg{Z}$ in the
neighborhood of $\theta \sim 3 \pi/8$. As explained in App.~\ref{app:chaotic}
this value of $\theta_h$ is in the ergodic regime, and $\avg Z$ is 0, which is
also what we obtain by direct numerical simulation.}.

We now attempt to reduce the cost of the numerical simulations by optimizing
the contraction ordering of the circuits simulated over $n=28$ qubits, see
Fig.~\ref{fig:q62}, and discarding every gate outside of the light cone of
$Z_{62}$ propagated backwards.  We look at both an ``open'' contraction, with
all output indices of the circuit uncontracted, and a ``closed'' contraction,
where a bit string is specified at the output of the circuit, typically leading
to a lower cost, although resulting in the computation of a single amplitude.
The optimized cost of  the open contraction is $2^{38.81}$, while for the
closed contraction we get $2^{38.45}$~\cite{footnote:FLOPs}.  This is to be
compared to $2^{39.13}$ (560 two-qubit gates over 28 qubits) for a full state
vector time evolution, similar to qsim's numerical simulations.  The similarity
between the three costs, which are exponential in $n$, is not surprising, given
that this circuit is in the high-depth regime, i.e., the depth of the circuit
is larger than the width of the lattice of qubits simulated. 

The above results suggest that the spontaneously broken phase, characterized by
non-zero quasi-static magnetization and finite correlation length, persists for
large values of $\theta_h$, corresponding to the regime of pre-thermalization,
see Sec.~\ref{sec:ergodic-1d} for details. In this case the state vector of the
system has overlap with the uniformly polarized initial state and at the same
time contains excitations that may not necessarily contribute to the local
magnetization. Most of the components of the state vector that do contribute to
the magnetization have correlation length smaller than the linear dimensions of
the qubit systems we considered in our simulations.

We emphasize that the situation here is remarkably different from the
out-of-time-order observable explained in Sec.~\ref{sec:otoc}. Indeed, the
effective volume of an out-of-time-order correlator between distant ancilla and
butterfly qubits can be an open cone with a diameter growing with the butterly
velocity. Consequently, the cost of tensor network contraction grows
exponentially with the square of the distance between the ancilla and butterfly
qubits, while the effective fidelity decreases exponentially with the cube of
this distance. In contrast, the effective volume of a local observable of a
physical system with a given correlation length is a cylinder with a diameter
proportional to this length. In this case the cost of tensor network
contraction grows exponentially with the square of the correlation length,
while the effective fidelity decreases exponentially with the cube of the
correlation length. 

\subsection{Non-local observable}

\begin{figure}
    \centering
    \includegraphics[width=\columnwidth]{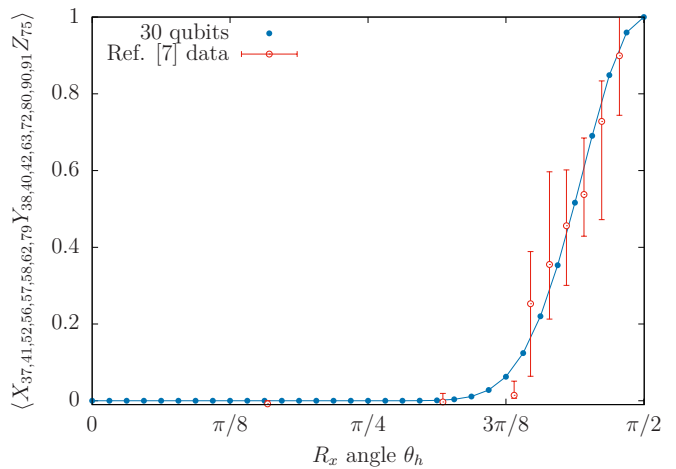}
    \caption{Expectation value as a function of $\theta_h$ for an observable
    composed of 17 Pauli operators, see Ref.~\cite{kim2023evidence} Fig.~4a,
    after 5 Floquet steps with Eq.~\eqref{eq:floquet} followed by single-qubit
    rotations. The numerical simulations agree with the experimental value from
    Ref.~\cite{kim2023evidence} Fig.~4a except for one point at $\theta_h$
    close to $3\pi/8$, where numerics are likely more accurate than the
    experiment. Each point per $\theta_h$ value is obtained in less than one
    second on an A100 GPU using the open source simulator qsim see
    Refs.~\cite{quantum_ai_team_and_collaborators_2020_4023103,
    isakov2021simulations, footnote:code}.} \label{fig:fig4a}
\end{figure}

Reference~\cite{kim2023evidence} also uses non-local Pauli observables of
support extending up to 17 qubits. Noting that circuits with $\theta_h = \pi/2$
are Clifford circuits, these observables are chosen by evolving a single qubit
$Z$ Pauli at time zero to obtain a high weight stabilizer at the final depth.
This immediately implies that the effective volume at $\theta_h \simeq \pi/2$
is an inverted cone between the high weight stabilizer at the final depth and
the single qubit chosen for the $Z$ stabilizer at the initial time. This
effective volume is smaller than the light cone, which grows from the final
non-local observable support to cover a larger number of qubits at the initial
time. This again is consistent with the relatively high effective fidelity
observed in Ref.~\cite{kim2023evidence}. For example, for the hardest non-local
observable presented in Ref.~\cite{kim2023evidence} Fig. 4a the effective
fidelity goes from $F_{\rm eff} \simeq 0.15$ up to $F_{\rm eff} \simeq 0.2$,
with an effective volume $V_{\rm eff} \lesssim 200$. We therefore expect, as
before, that the correct result can be well approximated with simulations using
less qubits than the light cone. 

Figure~\ref{fig:fig4a} shows the expectation value as a function of $\theta_h$
for an observable composed of 17 Pauli operators, see
Ref.~\cite{kim2023evidence} Fig.~4a, after 5 Floquet steps with
Eq.~\eqref{eq:floquet} followed by single-qubit rotations. The numerical
simulations agree with the experimental value from  Ref.~\cite{kim2023evidence}
Fig.~4a except for one point at $\theta_h$ close to $3\pi/8$. On similar
simulations where we can compare with exact results, such as
Figs.~\ref{fig:fig3b} and~\ref{fig:fig3c} in App.~\ref{app:more_numerics}, we
test that again we do not need to simulate the full light cone. Furthermore, we
note that in the same cases the experimental value in
Ref.~\cite{kim2023evidence} underestimates the exact value when it is close to
0. Therefore, the numerics are likely more accurate than the experiment for the
point where they do not agree. Each point is obtained in less than one second
per $\theta_h$ value on an A100 GPU using the open source simulator
qsim~\cite{quantum_ai_team_and_collaborators_2020_4023103,
isakov2021simulations, footnote:code}.

Similar to Section~\ref{sec:local_ferro}, we optimize tensor network
contraction costs for the circuit simulated with qsim with $n=30$ and
discarding gates laying outside of the observable's light cone.  For the open
contraction we get a cost of $2^{35.00}$, while for the closed contraction we
get $2^{17.44}$.  This is to be compared to $2^{39.24}$ for the full state
vector time evolution (161 two-qubit gates over 30 qubits), similar to qsim's
numerical simulations.  We now see a reduction in the contraction cost from a
state vector time evolution to an optimized open contraction (both exponential
in $n$), which is further reduced substantially by a factor of $~2\times 10^5$
for the closed contraction.  While the closed contraction results in only one
amplitude per computation, one can use it as a primitive to sample bit strings
from the circuit with a method such as that introduced in
Ref.~\cite{bravyi2022simulate} at a cost per sample that grows linearly in the
number of gates and in the closed contraction cost.  Samples can then be used
to estimate the expectation value of the observables studied.  These numerical
simulation costs are to be compared with those of random circuit sampling
experiments.  In the case of Ref.~\cite{morvan2023phase}, the hardest circuits
are estimated to require a cost larger than $2^{78}$.\\

In summary, a small effective quantum volume is the main reason why experiments
of Ref.~\cite{kim2023evidence} can be simulated on a classical computer. In
principle, a larger volume may be obtained by increasing the depth of the
circuits, hence increasing the spread of the observables, or by changing the
design of the circuits considered. On the one hand, without increasing the gate
fidelity, a potentially larger effective quantum volume will result in an
exponentially suppressed signal to noise ratio, which would make an experiment
impractical.  On the other hand, as in the case of the prethermalized
ferromagnetic states of Ref.~\cite{kim2023evidence}, the short correlation
length results in a small effective circuit volume, hence reducing the
classical computational cost.  In the end, an increased gate fidelity and a
careful design of the experiments is of utmost importance to reach the
beyond-classical regime.
\\

\textbf{Related works:} At the time of and after writing this manuscript, other
groups \cite{tindall2023efficient, beguvsic2023fast, anand2023classical,
liao2023simulation, patra2023efficient} have developed approximate simulations
of the experiments of Ref.~\cite{kim2023evidence} using different numerical
techniques.  For instance,
Refs.~\cite{tindall2023efficient}~and~\cite{patra2023efficient} use an
approximate evolution of the wave function, which is encoded as a compressed
tensor network that matches the connectivity of the experimental device.  Other
approaches are based on the approximate Heisenberg evolution of the
observable~\cite{beguvsic2023fast, anand2023classical, liao2023simulation,
patra2023efficient}.  These approximate evolutions are either carried out on a
truncated basis of Pauli strings~\cite{beguvsic2023fast}, on a compressed
matrix product operator representation of the
operator~\cite{anand2023classical}, or on a compressed two-dimensional tensor
network that matches the connectivity of the device~\cite{liao2023simulation}.
Interestingly, Ref.~\cite{beguvsic2023fast} also provides a hybrid method in
which both the operator and the wave function are represented with respective
tensor networks with device-like connectivity, each of which is partially
evolved.

Note that our approximation method is substantially different than the
aforementioned approaches.  More precisely, we introduce and exploit the
concept of an effective quantum volume to reduce the classical computational
cost.  Indeed, in the experiments of Ref.~\cite{kim2023evidence}, the effective
quantum volume is smaller than the maximum volume that can be classically
simulated.  Interestingly, the success of the approximate methods of
Refs.~\cite{tindall2023efficient, beguvsic2023fast, anand2023classical,
liao2023simulation, patra2023efficient} can be reduced to small quantum volume
for the experiments of Ref.~\cite{kim2023evidence} First, the same slow spread
of the observable under a Heisenberg evolution that gives rise to this small
effective quantum volume is responsible for the slow spread of correlations in
the evolved operator.  This, in turn, allows the Heisenberg evolution methods
studied by Refs.~\cite{beguvsic2023fast, anand2023classical,
liao2023simulation, patra2023efficient} to approximate well the results of the
experiments of Ref.~\cite{kim2023evidence} with low computational resources.
In addition, the sparsity of the qubit layout of these experiments, which is
key to the success of the tensor network methods mentioned
above~\cite{tindall2023efficient, beguvsic2023fast, anand2023classical,
liao2023simulation, patra2023efficient}, contributes to the slow spread of the
observables which is central to our work.
 
\section{Conclusion}

The computational cost of a local observable simulation can be upper bounded by
the effective circuit volume, that is, the minimal number of gates needed to
calculate this observable with a given precision. As was shown in
Ref.~\cite{mi2021information}, the effective volume can become prohibitively
large for classical simulations in the task of measuring OTOC for a random
circuit converging to the ergodic regime. The classical computational cost
increases as the dynamics approaches chaos, yet at the same time the value of
all local observables decreases even in the noise free case, and therefore the
precision required for a reliable measurement increases. 

In a realistic device gates have an error $\epsilon$, and often entangling
gates have the highest error compared to other operations. This further
decreases the signal-to-noise ratio. Consequently, noise also reduces the
effective volume, and the achievable classical computational cost that is
feasible with a given precision. Nevertheless, for sufficiently low noise, the
finite achievable effective volume would have a prohibitively high
computational cost. Ref.~\cite{mi2021information} is an example of a quantum
circuit experiment that explores this trade off. 

We would like to emphasize that the effective circuit volume strongly depends
on the specific observable. The effective volume is bounded in time and space
by the correlation time and length of the states that contribute to the
observable. For example, the prethermalized ferromagnetic states considered
above in the context of a Floquet transverse field Ising
model~\cite{kim2023evidence}  are characterized by static magnetization that at
first sight appears to be advantageous for the signal to computational cost
trade-off. However, such ordered states are characterized by a short
correlation length that limits the effective circuit volume and corresponding
computational cost even at long simulation times. In the ergodic regime, the
decay of time-ordered correlators (TOC) is characterized by very short
correlation time, that also limits the effective volume and computational cost. 

Summarizing the discussion above, in the quest for a beyond-classical regime in
noisy quantum processors, one has to choose observables carefully. The same
applies for comparing quantum simulations with brute force classical methods.
To illustrate this point, we presented  numerical simulations of a Floquet
circuit unitary experiment~\cite{kim2023evidence}, using generic methods, with
a computational cost of less than one second per data point on one GPU. 

We note in passing that the relationship between the effective circuit volume
and the noise parameters can be more complicated than the scaling $F_{\rm eff}
\sim e^{-\epsilon V_{\rm eff}}$ used above. Examples include systems that map
on free or weakly-interacting fermions and systems with conserved number of
particles. However, this does not change the fact that increasing noise reduces
the effective circuit volume and the magnitude of the expectation value
Eq.~(\ref{eq:P_S}), therefore reducing the cost of classical simulation.

\begin{acknowledgments}
We thank Orion Martin for writing the colabs that exemplify the simulations of
  Figs.~\ref{fig:fig4b} and~\ref{fig:fig4a}. We thank the Google Quantum AI
  team for numerous fruitful discussions. S.~Mandr\`a is partially supported by
  the Prime Contract No. 80ARC020D0010 with the NASA Ames Research Center and
  acknowledges funding from DARPA under IAA 8839. 
\end{acknowledgments}

\clearpage

\appendix
\section{Entanglement estimation for the circuits in
Ref.~\cite{mi2021information}}\label{app:otoc_entanglement}

\def\purity{\langle\!\langle {\rm Tr_{\mathcal{D}_1}} \rho^2 \rangle\!\rangle}

\begin{figure}[t!]
    \centering
    \includegraphics[width=0.5\textwidth]{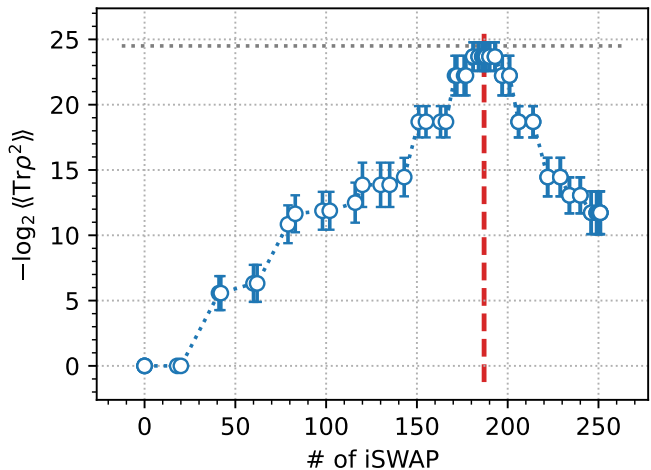}
    \caption{Averaged reduced purity $\purity$ for random OTOC Clifford
    circuits with a total number of $251$ entangling gates. The vertical dashed
    line corresponds to the number of entangling gates applied before applying
    the butterfly operator. The horizontal dotted line corresponds to the
    random Haar limit.}\label{fig:purity}
\end{figure}

Reference.~\cite{morvan2023phase} shows that the required bond dimension for
standard tensor network simulations of random circuits can be bounded by
computing the averaged reduced purity $\purity$. These circuits are similar to
the ones used in Ref.~\cite{mi2021information}.  Furthermore, the average
reduced purity is the same for an equivalent ensemble of Clifford circuits
(that is, using Clifford single-qubit rotations only). More precisely, we can
bound the required bond dimension $\chi$ as~\cite{morvan2023phase}
\begin{equation}
    \chi \geq \frac{F}{\purity},
\end{equation}
where $F$ is the target fidelity. Figure~\ref{fig:purity} shows the reduced
purity $\purity$ for random OTOC Clifford circuits with a total number of $251$
entangling gates, the largest OTOC circuits in Ref.~\cite{mi2021information}.
As expected, the lower bound for $\chi$ grows untill it reaches its maximum
when the butterfly operator is applied. At its maximum, $\chi$ almost reaches
its saturation value, meaning that there is no gain in using MPS or other
standard tensor network simulations for the largest circuits in
Ref.~\cite{mi2021information}.
 
\section{Chaotic dynamics with local time-ordered observable}\label{app:chaotic}

\subsection{2D lattice}
Many quantum simulation tasks in quantum chemistry, material science, and
physics require measurement of an expectation value of a local operator $o(t) =
\bra{\psi} U^\dagger O U \ket{\psi}$ with respect to some initial product wave
function $\ket{\psi}$ which can be prepared easily. Quantum dynamics in absence
of an extensive number of conservation laws exhibit quantum chaos at long time
scale. The long time asymptotic of a local observable scales as $o(t) \sim
1/2^{n/2}$ with the total Hilbert space $2^n$ of the $n$-qubit system.
Convergence to this limit is determined by the microscopic details of the
evolution operator $U$ and the locality/connectivity of the qubit array.
Nonetheless, the scaling with circuit depth $t$ can be deduced from the
following qualitative analysis of operator spreading. For example, consider a
single qubit Pauli $O$ located at the origin of a square 2D grid, evolving
under $U(t)$ without conserved quantities. In the Heisenberg picture $O(t)$ can
described using an expansion into Pauli strings, 
\begin{gather}
O(t) = \sum \alpha_{\upsilon} \prod_{(x,y)} \sigma_{\upsilon_{x, y}}, \label{eq:Pauli-exp}
\end{gather}
with $\upsilon_{x,y} = \{0,1,2,3\}$, and we assumed a two-dimensional square
grid of qubits labeled by pairs of indexes $x, y$. Note that the coefficients
are normalized as $\sum_{\upsilon}\left|\alpha_{\upsilon}\right|^2=1$. 

A single qubit Pauli operator $O(0)$ subject to a two-qubit entangling gate is
transformed into a superposition of two-qubit Pauli strings. Subsequent
application of two-qubit gates increases the size of the Pauli strings in the
superposition. This gives rise to operator spreading.  It proceeds in real
space such that at a given time $t$ all Paulis $\sigma_{\upsilon_{x y}}$
located sufficiently far from $O(0)$ (at the origin)  equal the identity, $I$.
Such dynamics can be characterized by a spatial distribution $w_O(t,r)$ of the
boundary of the support of $O(t)$, 
\begin{gather}
   w_O(t,r) = \sum_{\upsilon} \left|\alpha_{\upsilon}\right|^2 \delta(r -  r_\upsilon),
\end{gather}
where for a given Pauli string $\upsilon$, $r_\upsilon$ is the distance from
the origin of the farthest non-identity Pauli $\sigma_{\upsilon_{x, y}}$. In
the absence of conservation laws the operator spreading is ballistic. In these
cases  the distribution $ w_O(t,r)$ is sharply peaked around the boundary of
the support of the operator (a front). Then the time dependent size of the
operator can be described by the average,
\begin{gather}
    R(t) = \sum_{r} r\cdot w_O(t,r).
\end{gather}
Because of the approach to quantum chaos for $r<R(t)$ the coefficients
$\alpha_{\upsilon}$ are uniformly random such that the Paulis $B_{\upsilon_{x
y}} = \{I, X, Y, Z\}$ have equal probability with absolute values
$\left|\alpha_{\upsilon} \right|^2 \sim 1/4^{n_R(t)}$ where $n_R(t) \simeq \pi
R^2(t)$ is the number of qubits within the support. Ballistic propagation is
characterized by a butterfly velocity, $R(t) = v t$.

Consider an initial state $\ket{\psi}$ polarized along the $Z$-axis and
$O(0)=Z_{0,0}$ without loss of generality. In the
expansion~(\ref{eq:Pauli-exp}) only Pauli strings consisting of $Z$ and $I$
operators contribute non-zero values to the expectation value $o(t)$. In this
case 
\begin{gather}
    \bra{\psi} Z(t)\ket{\psi} = \sum_{\{\upsilon_{x,y}=I_{x,y},Z_{x,y}\}}
    \alpha_{\upsilon}(t) \sim 2^{-n_R(t)/2}.\label{eq:avZ}
\end{gather}
The latter estimate corresponds to the fact that the sum is over $2^{n_R(t)}$
terms of random signs. Including the ballistic spreading of the operator we
obtain,
\begin{gather}
     \bra{\psi} Z(t)\ket{\psi}  \sim 2^{-\frac{\pi}{2} v^2 t^2}.
\end{gather}

Note that the expectation value decays with a characteristics time scale $1/v$.
Using similar arguments we obtain the scaling of the correlation functions. For
example the equal time two-point correlator is
\begin{gather}
     \bra{\psi} Z_{0,0}(t) Z_{x,y}(t)\ket{\psi}  \sim 2^{-2n_R(t)+\Omega_{\rho }(t)},
\end{gather}
where, $\rho \equiv \sqrt{x^2+y^2}$, and,  $\Omega_{\rho}(t)$, is the area of
intersection of the supports of the operators $ Z_{0,0}(t)$ and $Z_{x,y}(t)$,
and $2n_R(t)$ is the sum of the support areas. The form of the function
$\Omega_{\rho}(t)$ is quite involved.  Similarly to the case of a single
observable (\ref{eq:avZ}) the leading factors in the dependence of
$\Omega_{\rho }(t)-2n_R(t)$ on $t$ contain  $(v t)^2$ and the dependence on
$\rho$ contains the  factor $\rho v t$.

The effective computational volume is given by the gates within a cone in
three-dimensional space-time. The base of the cone has area $n_R(t)$ and is
formed by qubits inside the support of $O(t)$. The height of the cone
corresponds to the circuit depth $t$. The circuit volume  $ V_{\rm eff} =
\frac{1}{3}n_R(t)t$ increases as $t^3$.

We note that the expectation value and the correlator decay rapidly on the time
scale $1/v$. Therefore the circuit depth and the respective circuit volume are
limited by the desired precision and the gate error via
Eq.~(\ref{eq:max-circ-volume}). 

In the presence of noise the observable equals,
\begin{gather}
    \textrm{tr}\left(\rho O\right) =  F(t) \times 2^{-\frac{\pi}{2} v^2 t^2}.
\end{gather}
Using the relation between fidelity $F(t)$ and the circuit volume $V_{\rm
eff}$, the depth $t$ of feasible quantum simulation is limited by
characteristic time $t<t_\delta$ that is the solution of the equation , 
\begin{align}
  \epsilon \frac{\pi}{3}\upsilon^2 t_{\delta }^3+\frac{\pi}{2} \left(\upsilon
  t_{\delta }\right)^2=\ln \left(\frac{1}{\delta }\right)\;.\label{eq:td}
\end{align} 
 
In the case of zero gate error 
\begin{equation*}
  t_{\delta }=\frac{1}{\upsilon }\sqrt{\frac{2}{\pi }\ln \left(\frac{1}{\delta }\right)}\;.
\end{equation*}\newline 
In the case of large gate errors the decay of the observable is determined by
the decay of the fidelity $F(t)$
\begin{equation}
  t_{\delta }=\left(\frac{3}{\pi \epsilon  \upsilon ^2}\ln \left(\delta
  ^{-1}\right)\right)^{1/3},\quad\ln \left(\frac{1}{\delta }\right)\gg
  \frac{\upsilon ^2}{\epsilon ^2}\;.\label{eq:tde}
\end{equation}\newline 
 
The cost of  tensor contraction, Eq.~\ref{eq:aeff}, depends on the butterfly
velocity. 
\begin{equation}
  {\rm cost}\sim 2^{2 \beta (v t)^2} \;,
\end{equation}
where $\beta$ depends on the type of the gate.

\subsection{1D and low degree 2D}~\label{sec:ergodic-1d}

It is instructive to also report the respective results for one-dimensional
system, a local observable in the interior of a long chain of qubits will scale
as,
\begin{gather}
    \bra{\psi} Z(t) \ket{\psi} \sim 2^{-\upsilon t}\;.\label{eq:z_1d_decay}
\end{gather}

\begin{figure}
    \centering
    \includegraphics[width=0.5\textwidth]{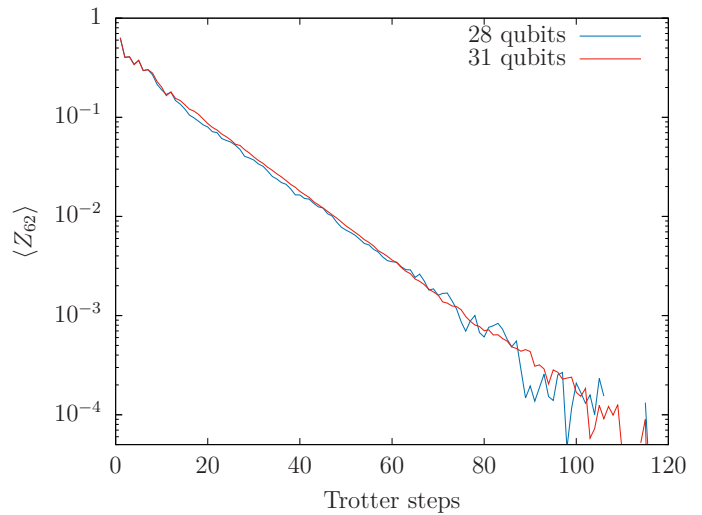}
    \caption{Exponential decay of the magnetization $Z$ for $\theta_h= 18 \pi/64$.}
    \label{fig:Z_ergodic_decay}
\end{figure}

\begin{figure}
    \centering
    \includegraphics[width=0.5\textwidth]{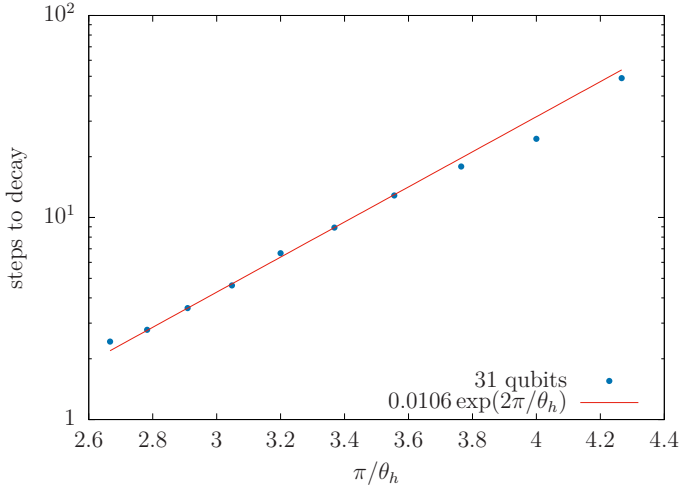}
    \caption{Exponential scaling of the magnetization steps-to-decay as a function of $\theta_h$.}
    \label{fig:decay_scaling}
\end{figure}

Note that in the case of experiment reported in Ref.~\cite{kim2023evidence} the
qubit lattice is substantially different from square lattice: the number of
nearest neighbors is $1$ for two qubits, $2$ for 89 qubits, and $3$ for 36
qubits. In other words this lattice is closer to one-dimension than to the
square lattice.

Figure~\ref{fig:Z_ergodic_decay} shows the exponential decay of the
magnetization $Z$ for $\theta_h= 18 \pi/64$ for the lattice in
Ref.~\cite{kim2023evidence}, in agreement with Eq.~\eqref{eq:z_1d_decay}.
Figure~\ref{fig:decay_scaling} shows the exponential scaling of the
magnetization steps-to-decay as a function of $\theta_h$. This is expected
scaling for prethermalization~\cite{PhysRevLett.115.256803}. For small
$\theta_h$,  deep in the prethermalization phase,  the  number of steps to
decay becomes very long because large phases of $ZZ$ gates make it hard to flip
the magnetizations of individual qubits.
 
\section{Additional numerical simulations}\label{app:more_numerics}
\subsection{Simulations for other figures in Ref.~\cite{kim2023evidence}}

\begin{figure}[t]
    \centering
    \includegraphics[width=0.5\textwidth]{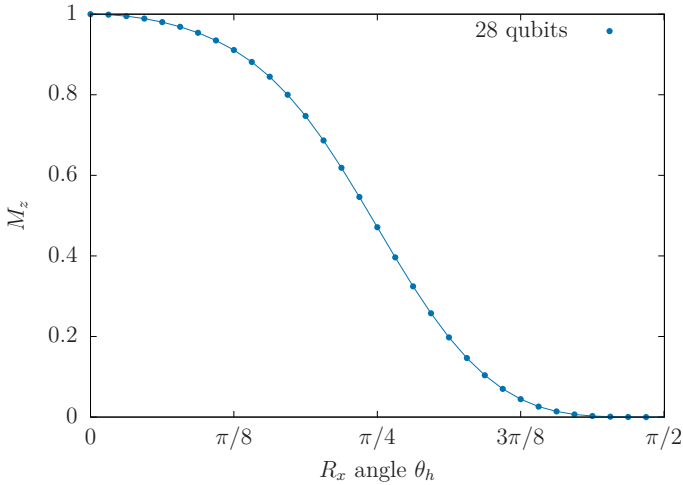}
    \caption{Average magnetization $\avg Z$ as a function of $\theta_h$ for the
    28 qubits of Fig.~\ref{fig:q62} after five steps of the Floquet circuit of
    Eq.~\eqref{eq:floquet}.}
    \label{fig:fig3a}
\end{figure}

\begin{figure}[t]
    \centering
    \includegraphics[width=0.5\textwidth]{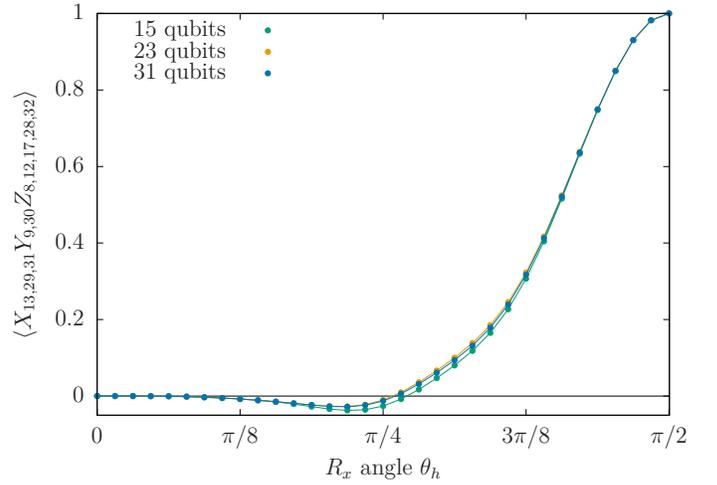}
    \caption{Expectation value as a function of $\theta_h$ for an observable
    composed of 10 Pauli operators, see Ref.~\cite{kim2023evidence} Fig. 3b,
    after 5 Floquet steps with Eq.~\eqref{eq:floquet}. The numerical
    simulations use 15, 23 and 31 qubits.}
    \label{fig:fig3b}
\end{figure}

\begin{figure}[t]
    \centering
    \includegraphics[width=0.5\textwidth]{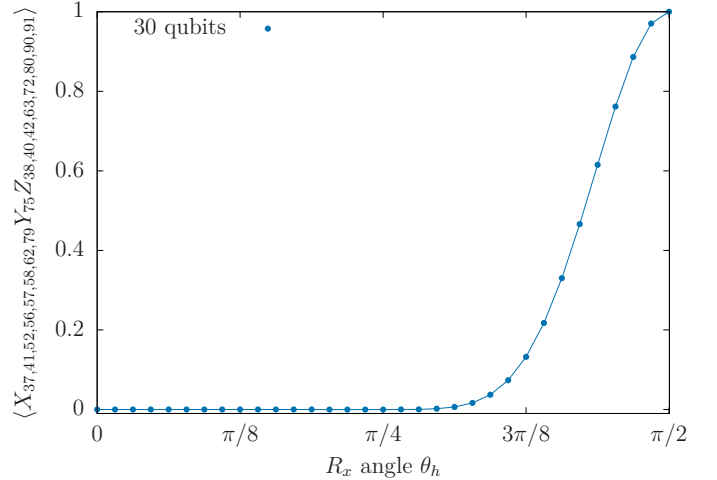}
    \caption{Expectation value as a function of $\theta_h$ for an observable
    composed of 17 Pauli operators, see Ref.~\cite{kim2023evidence} Fig. 3c,
    after 5 Floquet steps with Eq.~\eqref{eq:floquet}. The numerical simulation
    uses 30 qubits.}
    \label{fig:fig3c}
\end{figure}

For completeness, we include numerical simulations for the circuits in
Figs.~3a, 3b and 3c of Ref.~\cite{kim2023evidence}.  In all cases the numerical
simulations agree with the experimental results presented there.
Figure~\ref{fig:fig3a} shows the average magnetization $\avg Z$ as a function
of $\theta_h$ for the 28 qubits of Fig.~\ref{fig:q62} after five steps of the
Floquet circuit of Eq.~\eqref{eq:floquet}. Figures~\ref{fig:fig3b}
and~\ref{fig:fig3c} show the numerical simulation of an observable with 10 and
17 Pauli operators respectively, after five Floquet steps. In each case the
observable corresponds to a stabilizer of the $\theta_h = \pi/2$ Clifford
circuit obtained from the evolution of an initial single $\avg Z$ stabilizer.

Figure~\ref{fig:fig3b} compares simulations with 15, 23 and 31 qubits for the
10 Pauli operator measured in Ref.~\cite{kim2023evidence} Fig.~3b. Note that
there is a small difference between 15 and 23 qubits, although this difference
is smaller than the experimental error bars in Ref.~\cite{kim2023evidence}.
Nevertheless, there is good agreement between the simulations with 23 and 31
qubits. The light cone is 37 qubits, indicating that simulations of the smaller
effective volume are sufficient to obtain a good estimation of the observable. 

\subsection{Closed vs. open loops}\label{app:closed_vs_open}

\begin{figure}[t!]
    \centering
    \includegraphics[width=0.5\textwidth]{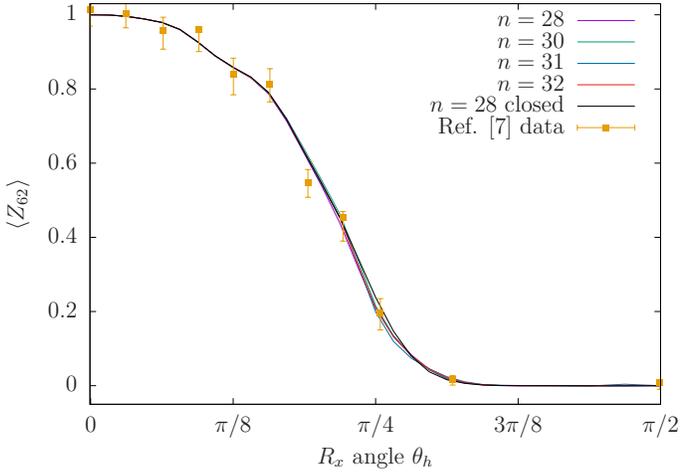}
    \caption{Numerical simulations for $\avg{Z}$ of the qubit labelled 62 in
    Fig.~\ref{fig:q62}, with 20 steps of the Floquet circuit of
    Eq.~\eqref{eq:floquet}. Figure also shows the experimental data reported in
    Fig.~4b of Ref.~\cite{kim2023evidence}.  We compare numerical simulations
    with close loops at $n=28$ qubits, see Fig.~\ref{fig:q62}, and simulations
    with $n=28$ up to $n=32$ qubits without closing the corresponding loops
    (open boundary).}\label{fig:fig4b2}
\end{figure}

Figure~\ref{fig:fig4b2} shows numerical simulations for $\avg{Z}$ of the qubit
labelled 62 in  Fig.~\ref{fig:q62}, with 20 steps of the Floquet circuit of
Eq.~\eqref{eq:floquet}. We compare numerical simulations with close loops, see
Fig.~\ref{fig:q62}, and simulations with $n=28$ up to $n=32$ qubits without
closing the corresponding loops (open boundary). We see a good agreement for
the estimation expectation value of $Z$ in all cases.  

\subsection{Convergence of $Z_{62}$ at 5 steps}\label{app:convergence}

\begin{figure}[t!]
    \centering
    \includegraphics[width=0.5\textwidth]{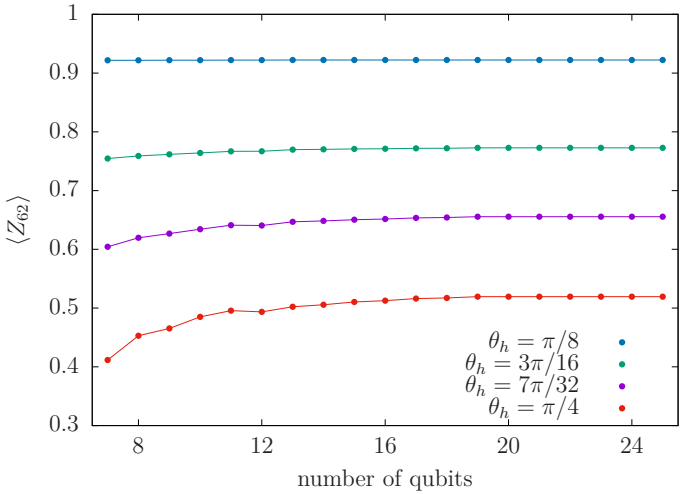}
    \caption{Numerical estimations of $Z$ at the qubit labeled 62 in
    Fig.~\ref{fig:q62} with the Floquet circuit of Eq.~\eqref{eq:floquet} and 5
    steps.}\label{fig:convergence}
\end{figure}

Figure~\ref{fig:convergence} shows numerical estimations of $Z$ at the qubit
labeled 62 in Fig.~\ref{fig:q62} with the Floquet circuit of
Eq.~\eqref{eq:floquet} and 5 steps. We show results for a growing number of
qubits, between 7 and 25, the size of the light cone. Smaller values of
$\theta_h$ have higher magnetization, and converge to the correct result with a
smaller number of qubits. As explained in Sec.~\ref{sec:local_ferro}, the
reason is that this corresponds to prethermalized ferromagnetic  states
characterized  by a short  correlation  length  that  limits  the  effective
circuit volume. Larger $\theta_h$ values presumably correspond to states with
larger correlation length.  Nevertheless, the magnetization converges to the
correct value with less qubits than the light cone even for the larger
$\theta_h = \pi/4$. Approximately after this value the state becomes ergodic
and the magnetization converges exponentially to 0, see App.~\ref{app:chaotic}.


\begin{thebibliography}{80}
\bibitem [{foo({\natexlab{a}})}]{footnote:FLOPs}\BibitemOpen
  \href@noop {} {}\bibinfo {note} {To translate into FLOPs for single precision
  complex scalars one should multiply by a factor of $8\times$.}\BibitemShut
  {Stop}
\end{thebibliography}
\end{document}